%% file: template.tex
\documentclass[journal]{vgtc}                

\usepackage{mathptmx}
\usepackage{graphicx}
\usepackage{svg}
\usepackage{times}
\usepackage{multirow}
\usepackage{enumitem}

\usepackage[bookmarks,backref=false,linkcolor=black]{hyperref} 
\hypersetup{
  pdfauthor = {},
  pdftitle = {},
  pdfsubject = {},
  pdfkeywords = {},
  colorlinks=true,
  linkcolor= black,
  citecolor= black,
  pageanchor=true,
  urlcolor = black,
  plainpages = false,
  linktocpage
}

\input{defs.tex}

\pdfoutput=1
\onlineid{0}

\vgtccategory{Research}

\vgtcinsertpkg

\title{Learning Vis Tools: Teaching Data Visualization Tutorials}

\author{Leo Yu-Ho Lo, Yao Ming, and Huamin Qu}
\authorfooter{
\item
 Leo Yu-Ho Lo, Yao Ming, and Huamin Qu are with the Hong Kong University of Science and Technology. E-mail: [yhload,ymingaa,huamin]@cse.ust.hk.
}

\shortauthortitle{Lo \MakeLowercase{\textit{et al.}}: Learning Vis Tools: Teaching Data Visualization Tutorials}

\abstract{Teaching and advocating data visualization are among the most important activities in the visualization community. With growing interest in data analysis from business and science professionals, data visualization courses attract students across different disciplines. However, comprehensive visualization training requires students to have a certain level of proficiency in programming, a requirement that imposes challenges on both teachers and students. With recent developments in visualization tools, we have managed to overcome these obstacles by teaching a wide range of visualization and supporting tools. Starting with GUI-based visualization tools and data analysis with Python, students put visualization knowledge into practice with increasing amounts of programming. At the end of the course, students can design and implement visualizations with D3 and other programming-based visualization tools. Throughout the course, we continuously collect student feedback and refine the teaching materials. This paper documents our teaching methods and considerations when designing the teaching materials.
} 


\CCScatlist{ 
 \CCScat{K.6.1}{Management of Computing and Information Systems}%
{Project and People Management}{Life Cycle};
 \CCScat{K.7.m}{The Computing Profession}{Miscellaneous}{Ethics}
}

\begin{document}


\firstsection{Introduction}

\maketitle

\input{content/fulltext_revising.tex}


\bibliographystyle{abbrv}
\bibliography{template}
\end{document}

%% file: defs.tex
\newcommand{\eg}{\textit{e.g.}}

\newcommand{\etal}{\textit{et al.}}

\usepackage[normalem]{ulem} 
\usepackage{xcolor}

%% file: content/fulltext_revising.tex
\abstract{Teaching and advocating data visualization are among the most important activities in the visualization community. With growing interest in data analysis from business and science professionals, data visualization courses attract students across different disciplines. However, comprehensive visualization training requires students to have a certain level of proficiency in programming, a requirement that imposes challenges on both teachers and students. With recent developments in visualization tools, we have managed to overcome these obstacles by teaching a wide range of visualization and supporting tools. Starting with GUI-based visualization tools and data analysis with Python, students put visualization knowledge into practice with increasing amounts of programming. At the end of the course, students can design and implement visualizations with D3 and other programming-based visualization tools. Throughout the course, we continuously collect student feedback and refine the teaching materials. This paper documents our teaching methods and considerations when designing the teaching materials.
}

Educating younger generations of visualization practitioners and researchers is a common and important duty in the visualization community. Sharing the experience of teaching visualization courses is a valuable way for course instructors to improve their teaching materials and help students achieve desired learning outcomes.

We offer an undergraduate data visualization course at HKUST every year. The course has two major learning outcomes: (1) understanding human perception and visualization design principles, and (2) applying learned knowledge to create visualizations with computer software. The first part is mainly covered in the lectures, while the second is taught by guided tutorials. Students are required to complete hands-on exercises and deliver a group project at the end of the semester.

In recent years, the demand for data analytics has grown rapidly, especially for business and scientific data analysis. As a result, our course enrollment includes students from diverse professions. In our latest class of 51 students, 30\% were studying majors outside of computer science and engineering (CSE), such as business, mathematics, and biotechnology. These students naturally have less programming experience than those with CSE backgrounds. \autoref{table:students} shows the statistics collected through a questionnaire on students' academic backgrounds and their relative programming experience. Such a diverse audience poses a big challenge to the teaching of guided tutorials.

In our previous data visualization course offerings, we mainly focused on teaching JavaScript libraries like D3 \cite{bostock2011d3} in tutorial sessions. However, students spent most of their time catching up with D3 programming concepts instead of putting the visualization knowledge into practice with the tools. We realized the limitation of this approach and took the opportunity to redesign the tutorials with a wider range of visualization tools. This has provided a better learning experience for both CSE and non-CSE students. This set of tools extends from D3 to include programming-based visualization libraries, such as Altair \cite{vanderplas2018altair} and Vega-Lite \cite{satyanarayan2016vega}, GUI-based visualization software, such as Tableau and Microsoft Excel, and web services provided by Google Colab \cite{colab} and Observable \cite{observable}. In this paper, we first present the design of our tutorial syllabus and materials, then we discuss the students' feedback and the lessons learned from redesigning the tutorials.

The tutorial materials are openly accessible on GitHub, released under the Creative Commons license.\footnote{\url{https://github.com/leoyuholo/learning-vis-tools}}

\begin{table*}[t]
\centering
\caption{Students' academic backgrounds and programming experience. We group students according to their school of study and further split engineering schools into CSE and non-CSE since CSE students have significantly more programming experience (rightmost columns). The middle columns show the programming language experience levels of the students. Notice that JavaScript is unpopular among all non-CSE students, and Python is the most common language in the whole class.}
\label{table:students}
\begin{tabular}{|l|rr|rr|rrr|}
  \hline
  &\multicolumn{2}{|c|}{\textbf{Class Composition}} & \multicolumn{2}{|c|}{\textbf{Experienced Languages}} & \multicolumn{3}{|c|}{\textbf{Year of Programming Experience}} \\
  \cline{2-3}\cline{4-5}\cline{6-8}
  \textbf{Profession } & \textbf{\# of Students} & \textbf{Percentage} & \textbf{JavaScript} & \textbf{Python} & \textbf{Less than 1} & \textbf{1 to 2} & \textbf{More than 2} \\
  \hline
  Business and Management & 6 & 12\% & 0 & 5 & 1 & 5 & 0 \\
  Engineering CSE & 34 & 70\% & 25 & 30 & 0 & 7 & 27 \\
  Engineering Non-CSE & 2 & 4\% & 0 & 2 & 1 & 1 & 0 \\
  Science & 7 & 14\% & 1 & 5 & 1 & 5 & 1 \\
  \hline
\end{tabular}
\vspace{-0.2in}
\end{table*}

\section{Related Work}

In the early years of teaching visualization courses, Hanrahan \cite{hanrahan2005} briefly introduced his undergraduate level visualization course, and Kerren \etal~\cite{kerren2008} gave a detailed discussion of their own experiences teaching visualization courses. They summarized the challenges of teaching a visualization course, of which the following relate to our concerns:

\vspace{-0.1in}
\begin{enumerate}[noitemsep]
  \item The diverse backgrounds of the audience, especially in programming knowledge;
  \item Providing hands-on experience to practice visualization knowledge;
  \item Choosing good datasets for practical exercises and to demonstrate visualization concepts and techniques; and
  \item Providing support for students to deliver quality projects.
\end{enumerate}
\vspace{-0.1in}

These challenges have been echoed by other researchers who have taught visualization courses at different institutions. Owen \etal~\cite{owen2013how} documented the authors' teaching experiences across eight different institutions. One of the authors structured a course that requires no programming, including tools like Microsoft Excel, Matlab and Geographic Information Systems (GIS). The authors also mentioned the necessity of teaching students the data preprocessing step before any subsequent visual presentations. Rohrdantz \etal~\cite{rohrdantz2014augmenting} documented the opportunities and pitfalls of using datasets from the Visual Analytics Science and Technology (VAST) Challenge for practical projects. However, not one course has demonstrated a complete solution in addressing the above challenges at once.

With growing interest in visual analysis, visualization courses are being offered by many institutions. Inspection of the publicly accessible teaching materials on the Internet shows that most instructors use D3 as the major focus for teaching visualization tools \cite{agrawala2018course, lex2018course, pfister2018course, scheidegger2018course}. This also includes the previous offering of our course. We realize the limitations of this approach and have initiated a rework of the materials to provide students a wider spectrum of visualization tools, teaching them to choose the most suitable tools for the particular task at hand.

Visualization tool development is undergoing rapid evolution. A number of key works pushing data visualization to a much wider audience include efforts by the research community on Altair, Vega-Lite, Vega \cite{satyanarayan2015reactive} and D3, commercial work on Tableau and Microsoft Excel; and the web services provided by Google Colab and Observable. We make good use of this wealth of improvements to build our tutorial content.

When creating teaching materials, we are inspired by the Coursera course of Daniel and Borg \cite{daniel2017coursera} on teaching Tableau. Another course on Coursera by Mannella \cite{mannella2017coursera} is the major reference for teaching Microsoft Excel. The book on data analysis by McKinney \cite{mckinney2017python} and his Pandas library \cite{mckinney2011pandas} are important resources for teaching data analysis with Python. The book by Murray \cite{murray2017interactive} is a comprehensive guide on D3. Finally, our course uses the textbook on visualization by Munzner \cite{munzner2014visualization} as a major reference book.

\section{Challenges and Requirements}

The challenges we spotted throughout the teaching process were: (1) the diverse backgrounds of the audience, (2) organizing materials coherently to suit the teaching schedule, and (3) picking good datasets for illustrating visualization concepts and conducting practical exercises.

\textbf{Challenge 1: Diverse Audience.} \autoref{table:students} shows the composition of the 2019 spring class, based on the 49 responses to our questionnaire. About 30\% of the students come from non-CSE professions, which echoes the academic background diversity mentioned in previous studies \cite{domik2011fostering, elmqvist2012leveraging, hearst2015vis, rushmeier2007revisiting}. Our course has a prerequisite of an introductory programming course, so every student has experience writing computer programs. It is worth mentioning the contrast between CSE students and non-CSE students regarding programming experience in JavaScript and Python. JavaScript became one of the most popular software development languages, so most of the CSE students have acquired skills in it. Python, however, has established its role in data analysis and scientific computing, so students who study these fields would choose Python over JavaScript. This implies that a change in the course syllabus is necessary to support the motivation of this group of students.

\textbf{Requirement 1: Tools with Minimal Setup.} One criterion in choosing tools for teaching is simple setup or even no setup at all. Complex environment setup creates problems that would discourage students from using the tools again after the course, especially those with less computing knowledge. Minimizing the cumbersomeness created by the setup also smooths out tutorial sessions, letting students focus on creating visualizations.

\textbf{Requirement 2: Tools Designed in a Visualization Language.} For the second criterion, we prefer software and libraries written in a visualization language (\eg, marks and encoding channels) that align with the visualization concepts and principles introduced in the lectures. This requirement favors tools built by the visualization community over those built by the statistics or data science communities.

\textbf{Challenge 2: Materials Organization and Teaching Schedule.} The role of the tutorials is to complement the lectures, provide a hands-on experience creating visualizations and, most importantly, prepare students to work on their course projects. Time is a key factor for delivering a quality course project \cite{kerren2008}. One of the goals in designing the syllabus is to help the students kickstart their projects as early as possible. We need to provide students the essential knowledge for getting started in the first few weeks.

\textbf{Requirement 3: Follow the Workflow of Data Science Projects.} We organize the materials according to the workflow of data science projects. A data project starts with exploring the dataset, then moves to forming hypotheses, evaluating them against the data, and progressively refining the solution. Similarly, we first teach students the tools suitable for data exploration, then tools for manipulating datasets to find more insights. Finally, we teach the tools for building a customized visualization solution for the specific problem at hand.

\textbf{Challenge 3: Choose Datasets for Demonstration.} Picking good datasets for demonstration is critical for students' engagement in the tutorials. Students are more active in learning when the datasets are relevant to their interests. For the best datasets, we have observed that students actively dig into different aspects of the dataset and communicate and share their findings.

\textbf{Requirement 4: Non-trivial, Manageably Sized, Real-World, and Engaging Datasets.} Throughout the course, we have iteratively refined the criteria of a good dataset for teaching: (1) It must have appropriate complexity to demonstrate visualization concepts in a non-trivial way. A good dataset also needs to convey the mission of demonstrating different visualization techniques or issues, such as visual clutter, dimensionality reduction, and data cleaning. (2) The size needs to be within a manageable range, otherwise, we need to provide a workaround. One of the datasets we use is very good in other aspects but large enough that we must make a workaround solution for it. (3) It must be a real-world dataset. (4) It must be relevant to the students' interests. We prefer real-world datasets over synthesized datasets because the stories found in real-world datasets are more authentic and stimulate students' curiosity to verify and explore the dataset or even go beyond the dataset to look for more facts on the Internet. This point is impossible for datasets synthesized for pedagogical purposes.

\begin{table*}[t]
\centering
\caption{Tutorial syllabus and datasets. Topics indicate the visualization tools and programming languages covered in the tutorials. Python-related tutorials are hosted on Google Colab, while JavaScript-related tutorials are hosted on Observable. Concepts may be repeatedly covered in different tutorials and demonstrated with different tools and datasets.}
\label{table:tutorial}
\begin{tabular}{|l|l|l|p{7cm}|}
  \hline
  \textbf{\#} & \textbf{Topic} & \textbf{Dataset(s)} & \textbf{Related Visualization Concepts/Techniques} \\
  \hline
  1 & Introduction and Microsoft Excel & University Ranking \cite{oneill2016} & Data Join, Long-form and Wide-form Data, Lie with Visualization, Visual Clutter \\
  \hline
  2 & Tableau & Superstore \cite{superstore2015} & Marks and Encoding Channels, Geographical Data, Diverging/Sequential Color Schemes \\
  \hline
  3 & Where to find visualizations and datasets? & & \\
  \hline
  4 & Python, Jupyter Notebook and scikit-learn & Pokemon \cite{rounakbanik2017} & Dimensionality Reduction \\
  \hline
  5 & Pandas and Altair & Spotify \cite{eduardo2018} & Data Cleaning, Data Join, Data Aggregation, Juxtaposition, Nominal/Quantitative/Temporal Data \\
  \hline
  6 & JavaScript and Observable & Pokemon & Marks and Encoding Channels \\
  \hline
  7 & Vega-Lite and Data Processing Libraries & Hong Kong Temperature & Data Preprocessing, Data Aggregation, Temporal Data, Heatmap \\
  \hline
  \multirow{2}{*}{8} & \multirow{2}{*}{SVG and D3} & Hong Kong Temperature & Grammar of Graphics \\
  \cline{3-4}
  & & Pokemon & Parallel Coordinates, Radar Chart, Glyph \\
  \hline
  \multirow{2}{*}{9} & \multirow{2}{*}{Interaction with D3} & Spotify (Top 10 Songs only) & Filtering, Missing Data Handling, Choropleth Map, GeoJSON and TopoJSON \\
  \cline{3-4}
  & & Pokemon & Interaction, Animation \\
  \hline
\end{tabular}
\vspace{-0.2in}
\end{table*}

\section{Tutorial Design}
\subsection{Course Background}

Our data visualization course at HKUST is a 14-week, project-oriented course. Students work in groups of three or four on the project, which makes up 50\% of the grade. Other assessments include in-class exercises, a presentation on a visualization of the students' own choice and a written exam. The class meets twice a week, with each session 1 hour and 20 minutes long, consisting of 60 minutes of lecture and 20 minutes for in-class exercises. The exercises take the form of criticizing a provided visualization or designing a visualization solution for a given scenario. At the beginning of the semester, lectures cover theories of human perception and their implications for visual designs. The topics then turn to visualization principles, different types of data and their case studies. Later in the semester, we focus our discussions on the evaluation of visualizations and several advanced topics covering recent research work.

In addition to the lectures, students have the chance to practice the learned knowledge with hands-on tutorial sessions. In each 50-minute session, students receive a set of tasks with guided instructions and demonstrations. These exercises are self-verifiable, so students can easily check the correctness themselves. Tutorial sessions only last for nine weeks; they are meant not to span the whole semester to leave time for students to work on their projects.

\subsection{Tutorial Syllabus and Materials}

The main role of the tutorials is to complement the lectures and provide hands-on experience using visualization tools. While visualization principles are introduced during the lectures, tutorials let students experience the principles in realistic cases, like how visual clutter appears when plotting a line chart with too many data points. There are also practical concepts that are hard to cover during lectures but are more suitable for teaching with practical examples. For example, long-form and wide-form data are very common but transforming from one to the other is a practical technique to learn. As emphasized by Kerren \etal~\cite{kerren2008}, \textit{learning by doing} is particularly important in visualization education.

Another mission of the tutorials is to prepare students to kickstart their projects early. We organize the syllabus to suit the data project workflow. The tutorials start with exploratory tools like Microsoft Excel and Tableau so students can quickly import data and make standard charts. We then provide pointers to where to find datasets and look for inspiration. We also encourage students to practice the learned exploratory skills with datasets in which they are interested and settle on a promising project idea. With a dataset in hand, we teach students data analysis tools to further investigate their datasets and find interesting stories hidden inside the data. In this part, we use Python with Jupyter Notebook \cite{kluyver2016jupyter} and supporting libraries like Pandas and Altair. These constitute a set of powerful tools for data analysis. Afterward, we lead students toward visualization libraries in JavaScript to equip them with the skills to tailor effective solutions for their projects.

\textbf{Tutorial 1: MS Excel.} We start tutorials with spreadsheet software because it is the most accessible visualization creation tool and widely used in different domains. This tutorial covers several fundamental techniques: (1) table lookup, which joins two datasets on a shared key, and (2) pivot tables, which can quickly rearrange tabular data for analysis. By introducing pivot tables, students also learn about transforming long-form data into wide-form and then using it as input for plotting charts in spreadsheet software. For the example dataset, we choose a dataset with rankings of different universities across recent years. The data is in two tables. The first has the year, the name of the university and its ranking. The second contains the mapping between universities and their corresponding countries. Students first join the two tables on university name, then plot line charts for different countries. We design the tasks to start with a country that has only a handful of universities, then move to those with over 50 universities, which creates visual clutter in the chart. Meanwhile, in the lectures, we introduce the \textit{lie factor} by analyzing a visualization on the press media \cite{friendly2000gallery}, which plots an up-ranking university with a downward curve; such a mistake will be reproduced in this tutorial if students do not flip the Y-axis on their line charts. All these experiences reinforce their understanding of the subject by doing the work themselves.

\textbf{Tutorial 2: Tableau.} As one of the most advanced GUI-based visualization tools, students can use Tableau to quickly draw visualizations without programming. The program also has an interface designed to work with the concepts taught in the lectures. For example, dimensions, measures, marks, and encoding channels are all commonly referred to when teaching visualization and subsequent tutorials on tools built by the visualization community. The demonstration dataset in this tutorial is several years of sales data of an international supplier. It is a pedagogical dataset widely used in teaching the functionalities of Tableau. However, students were not very interested in the dataset, which made learning less engaging. Compared to other datasets used in the tutorials, most of the attributes of this dataset are anonymized and students cannot relate to the numbers to form hypotheses and further examine the data. We will search for a replacement dataset for future offerings.

\textbf{Tutorial 3: Where to Find Visualizations and Datasets?} The third tutorial is a special one, designed to lead students into the visualization community on the Internet. At this stage, students have gone through the introductory part of the course and have the basic ability to analyze, critique and appreciate the visualizations they may see on the Internet. By introducing several popular communities and famous visualization authors, we show them where to look for inspiration and how to start picking their favorite visualization for the presentation assessment. In the meantime, we provide them a list of different data repositories to look for a dataset that is suitable for forming a project idea.

\begin{figure*}[t]
  \centering
  \scalebox{0.9}{
    \includegraphics[]{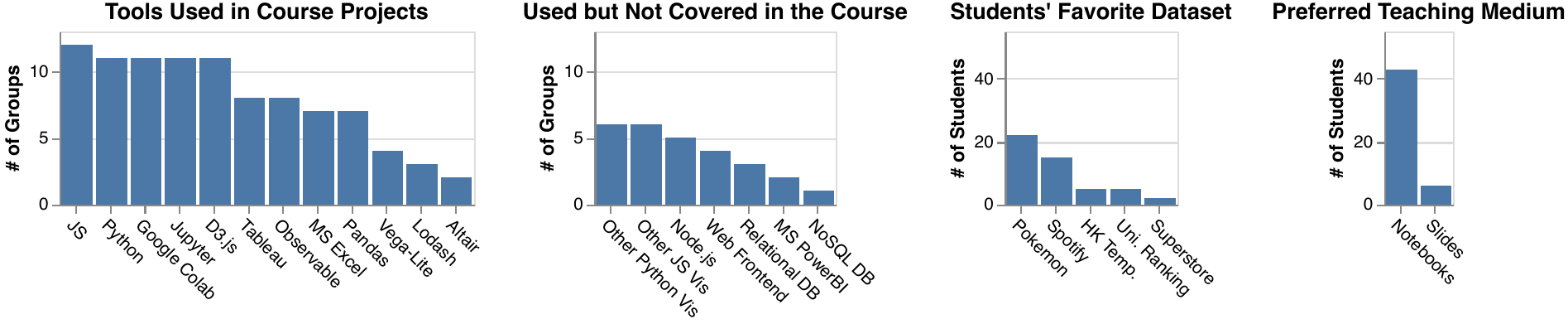}
  }
  \vspace{-0.1in}
  \caption{Results of 49 received questionnaire responses. From left to right: (a) visualization/supporting tools covered and students' usage in their projects; (b) visualization/supporting tools students used in their projects that are not covered in the course; (c) students' favorite datasets; and (d) students' preferences in using Jupyter/Observable notebooks or slides as the medium of teaching.}
  \vspace{-0.2in}
  \label{fig:content}
\end{figure*}

\textbf{Tutorial 4 and 6: Python and JavaScript.} Starting with the fourth tutorial, we focus on programming-based visualization tools, and from this point onward, we need to design our tutorials to cope with our diverse audience. For students coming from a non-CSE profession, we need to provide sufficient programming training, while for those who come from a CSE profession, we need to provide advanced materials to keep them engaged. Since we have planned to teach libraries in Python and JavaScript, we spend two sessions teaching the basics of each language. These two tutorials by no means cover all the important features in each language, so we decide to skip the language features not frequently used in the context of visualization creation. Examples are object-oriented programming features in both languages, iterators/generators in Python, and promises in JavaScript. Instead, we focus on the data types and how to use libraries to manipulate data. Students with prior experience in the languages can quickly finish these lab exercises \cite{brunner2016teaching}. To keep them engaged, we include additional tasks at the end of each tutorial, inviting students to explore the Pokemon dataset on their own. This Pokemon dataset attracts a large amount of interest from students, and we have observed them actively exploring the dataset and sharing findings with their peers. We simply provide the visualization tools and techniques they may need. For example, in tutorial 4, we guide students to perform dimensionality reduction with Python using the scikit-learn library \cite{pedregosa2011scikit}.

Traditionally, teaching the Python language involves either a command line interface or an integrated development environment (IDE); the development of Jupyter Notebook saves students the trouble. Because of its web-based interface, students can write their code with web browsers and run it on a remote server. Until recently, we have needed to help students set up such servers \cite{brunner2016teaching}. However, the online service Google Colab provides free access to its hosted solution so that students can start the tutorial directly without any setup or involving any command line. Not long ago, Observable began providing a similar platform for JavaScript, the difference being that the computation is done locally on the browser instead of on the remote server. These two services make the overall learning experience very smooth, and non-CSE students are more encouraged to bring visualization into their workplaces after the course.

\textbf{Tutorial 5 and 7 to 9: Pandas, Altair, Vega-Lite and D3.} Tutorial 5 revisits a couple of concepts practiced with GUI-based tools but using programming. Spotify's Worldwide Daily Song Ranking dataset is very suitable to teach techniques involved in data processing. It is a real-world dataset and contains missing values and unaligned time spans of some regions, so we can make use of its flaws to demonstrate data cleaning techniques. The only drawback of this dataset is its size: it is 369 MB and 45.2 MB when zipped, which takes a long time to download if the connection is overloaded. Fortunately, we managed to find a trick to make use of the infrastructure of Google Colab and its associated Google Drive service; the Jupyter Notebook running on Google Colab can directly download the dataset from Google Drive, and the whole process of loading the dataset only takes about 15 seconds. However, this does not apply to Observable, and later in tutorial 9, we have to limit the scope of the dataset to make it viable in JavaScript visualization.

For the visualization library, we pick Altair, which is developed by researchers in the visualization community, over a more popular plotting library, Matplotlib \cite{hunter2007} in the Python community. Altair's API is written in visualization language, the importance of which we have emphasized. Also, Altair is designed to integrate with Pandas, which largely reduces the friction of passing data across two different libraries. Lastly, it makes a seamless transition to the later tutorial on Vega-Lite. Because Vega-Lite is the underlying library of Altair, and itself is built on top of D3, moving down each layer requires students to do more programming while also provides increasing expressiveness for customized visualizations.

The tutorial materials are released under the Creative Commons license, while the datasets are licensed according to the original authors listed in the materials. We welcome any suggestions for improving or adapting them into any courses.

\section{Student Feedback}

We asked students to provide feedback by completing a questionnaire at the end of the semester. We received 49 responses and the results of which are shown in \autoref{fig:content}.

We are specifically interested in the tools students used in their projects. Compared to the previous offerings, instead of limited to D3, \autoref{fig:content}(a) shows that students have generally broadened their use of different tools, especially the use of Python, Jupyter/Observable notebooks and GUI-based tools. There are also tools that students actively learned outside the tutorials and included in their projects, like Matplotlib, plotly.js, Microsoft PowerBI, and supporting tools like web frontends/backends and databases, as shown in \autoref{fig:content}(b).

This is our first attempt to have most of the materials written in notebook format: Jupyter Notebook for Python and Observable Notebook for JavaScript. \autoref{fig:content}(c) shows very encouraging results, in that students are accepting this new form of learning materials.

Lastly, we received a lot of positive comments from students on both the tutorials and the course. In general, students liked the practical aspect of implementing the concepts learned from lectures and enjoyed exploring a wide range of visualization tools. Some advanced students even asked to cover more advanced techniques and provide optional practical exercises.

\section{Lessons Learned}

\begin{itemize}[noitemsep, leftmargin=0.15in]
  \setlength\itemsep{0.05em}
  \item Using Jupyter/Observable notebooks for teaching has several advantages over slides. Teachers can coherently interleave instructions with exercises in a single notebook. Students can try out new ideas on a new cell without overwriting previous results. This provides a safety net and encourages students to do more exploration. Several data science courses had already adopted Jupyter Notebook as the teaching medium \cite{brunner2016teaching, howard2019course, li2019course}, while Observable Notebook is more recent and dedicated to visualization communication. A recent visualization course by Heer \cite{heer2019course} is another pioneer adopting both notebooks in the teaching materials. We recommend more visualization courses to consider adopting notebooks for teaching.
  \item Students kickstarted their projects very early. In the presentation of the idea-forming phrase, several groups of students had already shown visualizations plotted with Tableau or used Jupyter Notebook to show their data analysis results.
  \item The disadvantage of lacking programming proficiency did not prevent students from delivering quality projects. In the course project presentation, we were impressed by a group of non-CSE students that presented a deep analysis on a dataset of their own domain expertise, which is difficult if not impossible when teaching only D3 in the course.
  \item While teaching a wide range of visualization tools, it is hard to cover each topic in depth. In compensation, we provide pointers to the materials of advanced topics and let the interested students explore, though these advanced materials may be hard to follow. In response to the student feedback, we consider to include optional materials in the future tutorials.
  \item Finally, datasets are the catalysts for student learning. We observed that students were more active in the class during tutorials 5 and 8, which use the Spotify and Pokemon datasets as examples. They investigated the streaming counts of different songs, tried to figure out the trends of their favorite songs in different regions in the world, and more importantly, shared their findings with their peers. The same thing happened with the Pokemon dataset as well. \autoref{fig:content}(d) shows that students liked these two datasets the most.
\end{itemize}

\section{Conclusion}

The landscape of visualization tools changes rapidly, and the tools used in the current offering may become obsolete within a couple of years. The interest of the audience may also vary from time to time and culture to culture. The course designers need to adapt the materials and teaching methods to their audience. Our work is merely an ingenious and expedient way to pull these resources together to address the challenges we have faced presenting to our students and provide them the experience of using these tools to create visualizations.

%% file: template.bbl
\begin{thebibliography}{10}

\bibitem{agrawala2018course}
M.~Agrawala.
\newblock {CS448B} visualization.
\newblock Computer Science, Stanford University, 2018.
\newblock Available: \url{https://magrawala.github.io/cs448b-fa18/} [Accessed:
  Jun 2, 2019].

\bibitem{rounakbanik2017}
R.~Banik.
\newblock The complete pokemon dataset, 2017.
\newblock Available: Kaggle Datasets,
  \url{https://www.kaggle.com/rounakbanik/pokemon}, Version 1, [Accessed: Jun
  2, 2019].

\bibitem{observable}
M.~Bostock, J.~Ashkenas, and T.~MacWright.
\newblock Observable.
\newblock Observable, Inc., 2019.
\newblock Available: \url{https://observablehq.com/} [Accessed: Jun 2, 2019].

\bibitem{bostock2011d3}
M.~Bostock, V.~Ogievetsky, and J.~Heer.
\newblock D3 data-driven documents.
\newblock {\em IEEE Transactions on Visualization and Computer Graphics},
  17(12):2301--2309, 2011.

\bibitem{brunner2016teaching}
R.~J. Brunner and E.~J. Kim.
\newblock Teaching data science.
\newblock {\em Procedia Computer Science}, 80:1947--1956, 2016.

\bibitem{domik2011fostering}
G.~Domik.
\newblock Fostering collaboration and self-motivated learning: Best practices
  in a one-semester visualization course.
\newblock {\em IEEE computer graphics and applications}, 32(1):87--91, 2011.

\bibitem{eduardo2018}
Eduardo.
\newblock Spotify's worldwide daily song ranking, 2018.
\newblock Available: Kaggle Datasets,
  \url{https://www.kaggle.com/edumucelli/spotifys-worldwide-daily-song-ranking},
  Version 3, [Accessed: Jun 2, 2019].

\bibitem{daniel2017coursera}
D.~Egger and J.~S. Borg.
\newblock Data visualization and communication with tableau.
\newblock {\em Coursera}, 2017.

\bibitem{elmqvist2012leveraging}
N.~Elmqvist and D.~S. Ebert.
\newblock Leveraging multidisciplinarity in a visual analytics graduate course.
\newblock {\em IEEE computer graphics and applications}, 32(3):84--87, 2012.

\bibitem{superstore2015}
Global superstore dataset, 2015.
\newblock Available: Tableau,
  \url{http://www.tableau.com/sites/default/files/training/global\_superstore.zip},
  [Accessed: Jun 2, 2019].

\bibitem{colab}
Google colab.
\newblock Google, Inc., 2019.
\newblock Available: \url{https://colab.research.google.com/} [Accessed: Jun 2,
  2019].

\bibitem{hanrahan2005}
P.~Hanrahan and K.-L. Ma.
\newblock Teaching visualization.
\newblock {\em Computer Graphics (ACM)}, 39:4--5, 2005.

\bibitem{hearst2015vis}
M.~A. Hearst, E.~Adar, R.~Kosara, T.~Munzner, J.~Schwabish, and B.~Shneiderman.
\newblock Vis, the next generation: Teaching across the researcher-practitioner
  gap.
\newblock {\em Panel at IEEE VIS}, 2015.

\bibitem{heer2019course}
J.~Heer.
\newblock {CSE512} data visualization (spring 2019).
\newblock Computer Science and Engineering, University of Washington, 2019.
\newblock Available:
  \url{https://courses.cs.washington.edu/courses/cse512/19sp/} [Accessed: Jun
  2, 2019].

\bibitem{howard2019course}
J.~Howard and R.~Thomas.
\newblock Practical deep learning for coders.
\newblock The Data Institute, University of San Francisco, 2018.
\newblock Available: \url{http://course18.fast.ai/} [Accessed: Jun 2, 2019].

\bibitem{hunter2007}
J.~D. Hunter.
\newblock Matplotlib: A 2d graphics environment.
\newblock {\em Computing in Science \& Engineering}, 9(3):90--95, 2007.

\bibitem{kerren2008}
A.~Kerren, J.~T. Stasko, and J.~Dykes.
\newblock Teaching information visualization.
\newblock In {\em Information Visualization: Human-Centered Issues and
  Perspectives}, pages 65--91. Springer Berlin Heidelberg, Berlin, Heidelberg,
  2008.

\bibitem{kluyver2016jupyter}
T.~Kluyver, B.~Ragan-Kelley, F.~P{\'e}rez, B.~E. Granger, M.~Bussonnier,
  J.~Frederic, K.~Kelley, J.~B. Hamrick, J.~Grout, S.~Corlay, et~al.
\newblock Jupyter notebooks-a publishing format for reproducible computational
  workflows.
\newblock In {\em ELPUB}, pages 87--90, 2016.

\bibitem{lex2018course}
A.~Lex.
\newblock {CS5630/CS6630} visualization for data science.
\newblock Computer Science, The University of Utah, 2018.
\newblock Available: \url{http://dataviscourse.net/2018/index.html} [Accessed:
  Jun 2, 2019].

\bibitem{li2019course}
F.-F. Li, J.~Johnson, and S.~Yeung.
\newblock {CS231n} convolutional neural networks for visual recognition.
\newblock Computer Science, Stanford University, 2019.
\newblock Available: \url{http://cs231n.stanford.edu/} [Accessed: Jun 2, 2019].

\bibitem{mannella2017coursera}
A.~Mannella.
\newblock Data visualization with advanced excel.
\newblock {\em Coursera}, 2017.

\bibitem{mckinney2011pandas}
W.~McKinney.
\newblock pandas: a foundational python library for data analysis and
  statistics.
\newblock {\em Python for High Performance and Scientific Computing}, 14, 2011.

\bibitem{mckinney2017python}
W.~McKinney.
\newblock {\em Python for Data Analysis: Data Wrangling with Pandas, NumPy, and
  IPython}.
\newblock O'Reilly Media, Inc., 2nd edition, 2017.

\bibitem{munzner2014visualization}
T.~Munzner.
\newblock {\em Visualization analysis and design}.
\newblock AK Peters/CRC Press, 2014.

\bibitem{murray2017interactive}
S.~Murray.
\newblock {\em Interactive data visualization for the web: an introduction to
  designing with D3}.
\newblock O'Reilly Media, Inc., 2nd edition, 2017.

\bibitem{oneill2016}
M.~O'Neill.
\newblock World university rankings, 2016.
\newblock Available: Kaggle Datasets,
  \url{https://www.kaggle.com/mylesoneill/world-university-rankings}, Version
  3, [Accessed: Jun 2, 2019].

\bibitem{owen2013how}
G.~S. Owen, G.~Domik, D.~S. Ebert, J.~Kohlhammer, H.~E. Rushmeier, B.~S.
  Santos, and D.~Weiskopf.
\newblock How visualization courses have changed over the past 10 years.
\newblock {\em IEEE Computer Graphics and Applications}, 33:14--19, 2013.

\bibitem{pedregosa2011scikit}
F.~Pedregosa, G.~Varoquaux, A.~Gramfort, V.~Michel, B.~Thirion, O.~Grisel,
  M.~Blondel, P.~Prettenhofer, R.~Weiss, V.~Dubourg, et~al.
\newblock Scikit-learn: Machine learning in python.
\newblock {\em Journal of machine learning research}, 12(Oct):2825--2830, 2011.

\bibitem{pfister2018course}
H.~Pfister.
\newblock {CS171} visualization.
\newblock Computer Science, Harvard University, 2018.
\newblock Available: \url{http://www.cs171.org/2018/} [Accessed: Jun 2, 2019].

\bibitem{rohrdantz2014augmenting}
C.~Rohrdantz, F.~Mansmann, C.~North, and D.~A. Keim.
\newblock Augmenting the educational curriculum with the visual analytics
  science and technology challenge: Opportunities and pitfalls.
\newblock {\em Information Visualization}, 13(4):313--325, 2014.

\bibitem{rushmeier2007revisiting}
H.~Rushmeier, J.~Dykes, J.~Dill, and P.~Yoon.
\newblock Revisiting the need for formal education in visualization.
\newblock {\em IEEE Computer Graphics and Applications}, 27(6):12--16, 2007.

\bibitem{satyanarayan2016vega}
A.~Satyanarayan, D.~Moritz, K.~Wongsuphasawat, and J.~Heer.
\newblock Vega-lite: A grammar of interactive graphics.
\newblock {\em IEEE transactions on visualization and computer graphics},
  23(1):341--350, 2016.

\bibitem{satyanarayan2015reactive}
A.~Satyanarayan, R.~Russell, J.~Hoffswell, and J.~Heer.
\newblock Reactive vega: A streaming dataflow architecture for declarative
  interactive visualization.
\newblock {\em IEEE transactions on visualization and computer graphics},
  22(1):659--668, 2015.

\bibitem{scheidegger2018course}
C.~Scheidegger.
\newblock {CS444} data visualization.
\newblock Computer Science, University of Arizona, 2018.
\newblock Available: \url{https://cscheid.net/courses/spr18/csc444/} [Accessed:
  Jun 2, 2019].

\bibitem{vanderplas2018altair}
J.~VanderPlas, B.~E. Granger, J.~Heer, D.~Moritz, K.~Wongsuphasawat,
  A.~Satyanarayan, E.~Lees, I.~Timofeev, B.~Welsh, and S.~Sievert.
\newblock Altair: Interactive statistical visualizations for python.
\newblock {\em The Journal of Open Source Software}, 3:1057, 2018.

\bibitem{friendly2000gallery}
Why does college have to cost so much?
\newblock Ithaca Times, 2000.
\newblock Available: Gallery of data visualization,
  \url{http://euclid.psych.yorku.ca/SCS/Gallery/context.html} [Accessed: Jun 2,
  2019].

\end{thebibliography}
